# Computability vs. Nondeterministic and P vs. NP

Zhou, Jian-Ming  (email to: yu.li@u-picardie.fr )
May 2013

**Abstract**: This paper demonstrates the relativity of Computability and Nondeterministic; the *nondeterministic* is just Turing's *undecidable Decision* rather than the *Nondeterministic Polynomial time*.  Based on analysis about TM, UM, DTM, NTM, Turing Reducible, β-reduction, P-reducible, isomorph, tautology, semi-decidable, checking relation, the oracle and NP-completeness, etc., it reinterprets The Church-Turing Thesis that is equivalent of the Polynomial time and actual time; it redefines the NTM based on its undecidable set of its internal state. It comes to the conclusions: The P-reducible is misdirected from the Turing Reducible with its oracle; The NP-completeness is a reversal to The Church-Turing Thesis; The Cook-Levin theorem is an equipollent of two uncertains. This paper brings forth new concepts: NP (nondeterministic problem) and NP-algorithm (defined as the optimal algorithm to get the best fit approximation value of NP). P versus NP is the relativity of Computability and Nondeterministic, P≠NP. The NP-algorithm is effective approximate way to NP by TM.

**Keywords**: reducible undecidable actual-time λ-calculus isomorph tautology oracle TM NP NPC

**Contents**:
 1. The Turing's Machine: Halting or fail
 2. The Church-Turing Thesis: Actual Time and Polynomial Time
 3. The λ-calculus and β-reduction
 4. Nondeterministic: DTM, NTM and TM
 5. The P-reduction and Isomorphism
 6. The P-reduction and Semi-decidable
 7. The Oracle and NP-completeness
 8. The NP Theory: NP ≠ P

Up till now, the "Turing Machine" is still somewhat mysterious although this concept has been permeating almost sense of cyberculture, the definition of Turing Machine whether its Informal or formal descriptions are clear and seems pellucid, however has been near-universally accepted. The TM based on intuition simulation instead of an axiom or essential principle, people are easy to get along with its demonstrating, but hard to explore its profound mystery by one's own initiative, for example, what to say or how to say the Church Turing Thesis as a hypothesis?

In the last century when mathematicians awaked the "Mathematics: the Loss of Certainty" (M. Kline 1908-1992), K. Gödel (1906-1978), A. Church (1903 - 1995), A. Turing (1912-1954), etc., attended the identification of computability with the algorithm, which involves chiefly the idea of *Nondeterministic,* instead of the "incomputability", in fact people always avoid to use like this concept in that paradox may contain. But the famous Turing's *Undecidable is often not comprehended as* the concept *Nondeterministic* rather has been arrogated by the prevalent "NP problem", which become a millennium problem indeed.

**1. The Turing's Machine: Halting or fail**

A. M. Turing had demonstrated many notions and definitions in his 1936 paper "On Computable Numbers, with an Application to the Entscheidungsproblem" [2] as follows:

The a-machines and c-machine:
"If at each stage the motion of a machine is completely determined by the configuration, we shall call the machine an 'automatic machine' (or a-machine). For some purposes we might use machines (choice machines or c-machines) whose motion is only partially determined by the configuration." [2] (§2)

The computing machine of a-machine:
"If an a-machine prints two kinds of symbols, of which the first kind (called figures) consists entirely of 0 and 1 (the others being called symbols of the second kind), then the machine will be called a computing machine." [2] (§2)

The circular machine and circular-free machine:

"If a computing machine never writes down more than a finite number of symbols of the first kind it will be called circular. Otherwise it is said to be circle-free. "

Please note the word *circular* used here is aimed at the computable sequence, and then the circular machine differs from a machine in Circular:

"A machine will be circular if it reaches a configuration from which there is no possible move, or if it goes on moving, and possibly printing symbols of the second kind, but cannot print any more symbols of the first kind. The significance of the term 'circular' will be explained in §8."

Turing strictly distinguish the computable number and computable sequence, in fact the former is the number as a computer and the latter is the reclusive enumerable (*computably enumerable — c.e*.) set of computes. The computable sequence and computable number were distinguished:

"A sequence is said to be computable if it can be computed by a circle-free machine (here should be apprehended as a machine in circular-free —added). A number is computable if it differs by an integer from the number computed by a circle-free machine."[2] (§2) (namely, the computable number can be computed by circular machine, —added).

These definitions of complex classification are based on different hierarchies, roughly, the *print*, *write*, *output* (on tape) corresponds the computable number that is also the circular machine that can prints only "1s" and "0s", for example prints infinite "0", attention please, this circular machine is running in circular-free at the moment (Turing as a crackerjack cryptographist knows too much the *state* of machine); thus the *computable sequence* can be computed by a circular machine in circular-free state. Therefore the *circular* and *circular-free* are either the machine (number consists of "1s" and "0s" ) or the state of machine itself, which is very different in Turing's proof. The *circular-free* number (for example the infinite non-decimal is incomputable irrational number) is difference from the *circular* number, a circular number (as computable machine) in circular-free is computable for the circular number, but unable (in circular) for circular-free number. There "gives a certain insight into the significance of the idea 'circle-free'" in Turing's proof, demonstrated the complex relations between the algorithm and number, which on the state of machine itself in the *actual time* [15] is different from the abstract diagonal process.

Turing thought the diagonal process may represent as judgment (Entscheidungsproblem):

"… the problem of enumerating computable sequences is equivalent to the problem of <u>finding out</u> whether a given number *is* the D.N of a circle-free machine, and we have no general process for doing this in a finite number of steps. In fact, by applying the diagonal process argument correctly, we can show that there cannot be any such general process."

"The simplest and most direct proof of this is by showing that, if this general process exists, then there is a machine which computes J. This proof, although perfectly sound, has the disadvantage that it may leave the reader with a feeling that 'there must be something wrong'. The proof which I shall give has not this disadvantage, and gives a certain insight into the significance of the idea "circle-free". It depends not on constructing J, but on constructing J ', whose n-th figure is Yn(n)."[2] (§8, subline added)

Turing's key is that "It depends not on constructing J, but on constructing J ' ", in his paper, the "J" is circular machine and "J'" the circular-free machine. Actually, the *constructing J'* is a judgment pseudo mechanical process that may be called *Turing's UM* (UTM), which is different from the Universal Machine (UTM or UM) via simulation:

"It is possible to invent a single machine which can be used to compute any computable sequence. If this machine I is supplied with a tape on the beginning of which is written the S.D of some computing machine M, then I will compute the same sequence as M." [2] (§6)

A circular machine is able to compute a circular number, or decide (yes or no) a circular-free number by own state (in circular-free or circular), but the circular-free machine is untenable. For accurately discussing going on we will redefine some concepts in this paper:
1. We should distinguish the *judgment* from *decidable*, a judgment is decidable if its decision yielded by an TM otherwise undecidable.
2. We will use respectively two notions of *Halting* and *fail*, the *halting* references the work state of a circular machine itself running in circular-free. The *acceptable* state or *reject* for languages are the halting. For example, a TM yields inputs of computable sequence and stop, or of infinite recurring decimal, for instance an infinite "0s"; or a judgment for language is decidable while a TM working in halting state.
3. The "*halting problem*" is that whether the state of TM in halting state or "not halting" (*fail*).
4. The *fail* is "running" in dead loop of machine, namely the machine is untenable
5. The common word "stop" is ambiguous, that may be halting while a TM is in output state, or cracked down in dead loop namely *fail*. Besides hardware malfunction is "stop".
6. Especially, according Turing's connotation here is reluctantly *circular-free machine* —not circular machine, not compute machine, not a-machine —rather the c-machine — *partially determined by the configuration*, which is just the famous Halting Problem. The *decidable* of Judgment is computable, but the *Undecidable Decision* is *undeterminable* of judgment— the *Nondeterministic* of computability itself, its opposite not incomputable rather Turing's *oracle (*omniscience).

The UM used in Turing' proof (TUM) is to attempt to construct a (deterministic) machine in circular state to compute (decide) a circular-free number and that is impossible:

"…Thus both <u>verdicts</u> are impossible and we conclude that there can be no machine D.

…Now let F be a machine which, when supplied with the S.D of M, will write down successively the S.D of M, of M1, of M2, … (there is such a machine). We combine F with C and obtain a new machine, G. In the motion of G first F is used to write down the S.D of M, and then C tests it, :0: is written if it is found that M never prints 0; then F writes the S.D of M1 and this is tested, :0: being printed if and only if M1 never prints 0; and so on. <u>Now let us test</u> G with C<u>.</u> If it is found that G never prints 0, then M prints 0 infinitely often; if G prints 0 sometimes, then M does not print 0 infinitely often.
Similarly there is a general process for determining whether M prints 1 infinitely often. <u>By a combination of these processes we have a process</u> for determining whether M prints infinity of figures, i.e. we have a process for determining whether M is circle-free. <u>There can therefore be no machine R.</u> "[2] (§8) (subline added)

In fact the machine D is a c-machine that partially determined by the configuration and partially by human, which is actually the judgment (Entscheidungsproblem). The verdict "the Halting problem is undecidable" is declared by human via untenable of TUM: *There can therefore be no machine G*.

We may make out the *Halting problem* is *pseudo* TUM to simulate a computation to computations. The *halting* is a judgment *value* rather a computable number, although the both are often represented with same symbols "0" and "1". Whether the both are equal in actual time simultaneously that is nondeterministic.

This differentiate of Halting from Stop can help us to make some difficult definition much clearer. For example, S. Cook redefined Turing's Hating problem as "HP = {‹M› | M is a Turing machine that halts on input ‹M› } "[6], the *halts* here is the state of *accept* or *reject* to ‹M› rather than the *fail* in the M itself, then the HP here is decidable rather than untenable, the *fail* often is specially the *Halting problem* that is distinguishes form "halt" or "stop".

The words "simulant" and "pseudo" may correspond "effective" and "undecidable", the TM is effective for any computable problem but untenable (*fail*) for *halting problem*, the *Undecidable Decision* is gained by pseudo judgment rather than yielded by simulant computer, the Entscheidungsproblem is Judgment that may partially by computation (*halting*), but *the halting problem* (all judgment) is untenable for TM. Whether every Judgment can decide by TM or any judgment can decide by a computer that is nondeterministic.

Two key points:
1. The TM is Decidable for computable numbers or sequence, in which the *computable* is just decidable decision, that may notates by P algorithm = P problem for TM.
2. Complete expression: the Halting Problem is Undecidable Decision by TM.

The Halting Problem often describe as that: let a TM to decide the TM itself, which is *pseudo* UM rather *simulant* as TM. In fact the TM is equivalent with computability whether it stop or not (the not stop may be a computable for example the infinite circle number), if the UM is judgment that must *stop* (execute a decision) to input a value (yes or no) otherwise it is unable for the judgment, which is just the Halting Problem.

## 2. The Church-Turing Thesis: Actual Time and Polynomial Time

The computation is primitively the human's behavior in algorithmically operation, intuitionally, —a kind of effective and passive operation—so called the *mechanical steps*, the *effective* means that must be able to gain of deterministic numbers or result and the *passive* means that is controlled by certain rule. The concept of algorithm emphases the rule and often glossed over the actual time as operator self, "The algorithm is the purest form in all knowledge theories, which in certain significance may be regarded the final representation of *instrumental rationality* (Max Weber's noted "*rationally pursued and calculated*"). In philosophy significance, the ultimate problem about the knowledge and the theories about it, all represent finally in the algorithm theory. The algorithm and computer theory have uncovered adequately that problems can be solved via computation all are solved finally by mechanical steps (the Church-Turing's thesis). In this meaning, the notion of algorithm is just the computation in mechanical steps. In fact the symbol *1* and *0* in contemporary computer theories represent the pure *steps*, that is but the classical *mechanical steps* has been replaced by this pure algorithm symbol. The 1 and 0 as the relation of pure symbol is the step's relation. (The 1 in algorithm theory is the true *nature number —true number*, common digital number is but a human's convenience form of the true number 1.) The concept of mechanical steps connotes the meanings of linear process in actual time. The Universe in classical significance is the unique (absolute) time; the actual time has no parallelism. Therefore the algorithm of computability possessed the linearity essence, which is the final meaning of the concept algorithm that possesses the computability. Of course the issue is not so simple and pure, the problems can be solved by linear computation are a very few in our problem world, which is far from whole algorithm theory (computability theory). The true profound problem in algorithm theory is the property of the *computability* itself, which examines further minutely the property of *incomputability*, which In philosophy meaning is the problem how they (computability and incomputability) constricts our world. The essence of algorithm (computability) and incomputability is only explored by philosophy, but which concretely represents in the computation theory, especially contemporary algorithm theory based on the computer theory" (Zhou: The Algorithm theory and Chinese Reason[13])

So called the idea of classical time is intuitional even and linear time process. The algorithm as ancient mathematics had abstracted actual time. The word *calculate* was from Latin calculus, calx is small stone for gaming or used in reckoning, that become the *number* after to be enumerated in actual time— "one" per "one time", that is the very primitive significance of the *effective*. Intuitionally the *effective* procedure is *actual* process controlled by rule, for example, to count from 1 to 100, count down, reckon with zero, clearing of accounts, I lost count…, these practical behaviors are executing actually namely in the actual time [1].

The human as it were is the prima algorithm, for example, to number, to count is positively steps in format operation and theoreticians termed as *enumerate* (steps); *in* *fact* the Nature Number as the basic enumerate operation is also the primitive algorithm. The research of recursive function especially the primitive recursive function reveals the relation of function and algorithm, which has same computable facility degree in actual time, that is the concept of *effectively computable function*, in which the function as the relation between functions and as algorithm in actual time are identifying.

In intuitional meaning the effectiveness is based on the actual time namely that has same facility degree in actual time, and the computable connotes the facility of algorithm, the Church-Turing Thesis as hypothesis stands this facility equipollence.

Intuitionally the notions of effective computable means a means carry out in finite steps and must be able to get result, which in mathematics is the *Polynomial Time* that replaced the length of time with the length of symbol strings. Even more important, the polynomial time labels a linear parallelism— the number of computational steps grows as a power of the number of input problem —the relative adaptability of capability, which is just the *computable* or *computability*. Computability is computable in actual time, which is an adaptable relation between computation and problem.

The contemporary computer is based on electro- logic- circuits and electronical pulses acting that are represented as the symbol of "0" and "1" service as the *mechanical steps*. In fact all operations of computation is finely accomplishing in the addition in CPU, the frequency of CPU is a mark of computer's physics capability, that is to say the frequency only shares the unique actual time. That is to say all real computers are running in the actual time namely, the contemporary computer is an authenticity of actual time, or speaking the actual time is this general identity in all computation means, which is also the meaning of the Church-Turing Thesis.

The symbol strings of theory are represented ultimately by the symbol of "0s"and "1s", the length of which is called the *polynomial time* that is the mathematical form of the actual time. Hence one can see the consistence between the Church-Turing Thesis in the Polynomial Time and in the actual time.

The TM imitated human's calculates behavior has successfully represented in symbolization of configuration which is a counterpoint through the read-write head between the internal state and tape, chiefly the former, so called "n-tuples" (quadruple, quintuple, tabular, septuple and so on), which represents the combination mode of the abstract form as algorithm rule and computation process as actual time, namely the mechanical process of symbol or symbolization of mechanical steps. The configurations (mechanical or symbolic) running in itself state that is in actual time, for example the tape and storage, or the heard and transitional function. In fact all effective computable means (computability) is equal in actual time, the famous Church-Turing thesis stands up for this complex identical idea, which directly represents the identification of recursion, the λ-calculus, the Turing machine, and general computability, these transcendent identity upon actual time can't be proven only regarded as a kind of hypothesis, but we can still explain their mechanism.

The symbolization of the actual time is unfolding along another way that is the so called the Polynomial Time. The polynomial time is polynomial expression roughly the length of strings, which is not symbolization of actual time rather the capability of computation in actual time. The Polynomial Time often is used as "in (or bound by) polynomial time", which is a comparing of relative capability, for example, an algorithm is said to be of polynomial time that references its running time is upper bounded by a polynomial expression in the size of the input in TM. This relative capability has been intuitionally accepted as the synonym of "feasible" or "efficient".

In another hand the TM can represented a string by re-coding, the string of TM including its input is just the UTM, which means the TM is equivalent with the UTM in the form of polynomial time. The symbolizations of TM (in actual time) by the strings is equivalent with algorithm in polynomial time. An algorithm in polynomial time is equivalent with a-automaton in actual time (namely *a circular machine in circular-free state*, see above section1.; The relation of string or symbol and mechanical steps is another deeper topic that is not discussed here) .

This equivalence stood by the Church-Turing thesis, namely all effective computable algorithm is computable of TM. In another word, so called bounding in Polynomial Time is same with the usage "in actual time, owing to this transcendental property could not be explained The Church-Turing Thesis only serve as a hypothesis.

The TM in actual time including all practice computers, that is to say all practice computes or all algorithms of polynomial time are equivalent in computability rather than in practical running time, the concept of computation complexity involves latter often.

### 3. The λ-calculus and β-reduction

The algorithm as the rule has often ignored the actual time, the λ-calculus (by A. Church 1930[2]) is a formalization of the algorithm as human's abstract operation in actual time. The λ-calculus is the yielding of function rather than function relations, the latter relations are always to be linked with symbol "=", but the λ-calculus is a kind of formula of yielding or abstracting out functions rather than invoking them, thereby does not concern these function (anonymous) rather the abstraction itself. The λ-calculus is abstract manipulation of special symbols, which integrates all operating in a string of symbols: ".", variable, function, and parenthesis after the "λ", for instance λ x.y, in which the free variable x is transformed a parameter of domains in y, the free variable y became an anonymous function "Y", namely f(y) or f(x), this operation is commonly called *bounding* or *abstracting*, for example:

λ x.x   free variable x is be bounded by x, namely x = x (or y = y, so called α-conversion) , which sets up the identity function;

λ x.X   free variable x is be bounded by function f(x), namely X=x, or f(y) = y (x or y is identity variable in domain and codomain of f(x) or f(y)), which sets up a constant function.

Similarly:

λ x.(x × x) yields: f(x)=x × x, multiplication function.

The λ-calculus ought to be understood operationally instead of logically, which could be regarded as mathematization of logic deduction. In actual time the manipulation of λ-calculus only is one time in any time, thereby only a single variable each time is executable in procedures of a λ-calculus, and the multi-variable must be calculated in *curry*:

λ x.(x+y)  executes:
λ x.(λy.(x+y))

   An application of λ-calculus defined formula: (λx.y)z, which is executes as β-reduction, to unfold two procedures:
First to bound x in function f(y);
Second substitute parameter y in f(x) with z.
People usually use additional symbols to unfold the *β-reduction*, commonly introduced operation formula like the formula Y[x;=z] that executes to substitute x with y in Y, namely an application of (λx.y)z may be reduced as Y[x:=z], commonly fetching in a new symbol ⊢ to express this reduction, namely introduced symbol "⊢" between the two procedures:
(λx. Y)z ⊢ Y[x:=y].

   For example, (λn.n×2)3, first operation yields function f(x) = x × 2; second operation executes (x × 2)[x:=3] and yields f(2) = 3×2.

   A λ-calculus application may be infinite, for example
(λx.xxy)( λx.xxy) ⊢ (xxy)[ x:= λx.xxy)]:
(λx.xxy)( λx.xxy)y
(λx.xxy)( λx.xxy)yy

……,

That may correspond a Turing's *circle computable sequence*.  In fact the λ-calculus similar TM that can compute computable number or sequence, Turing has proved that the λ-definability function is λ-K-definability [3].  All recursive functions are λ-calculus, and all λ-calculus numeric functions are recursive functions.

The free variable and bounding are important concepts in computability theory. A variable x is free if {x} for all x, A variable x in Y is bound if x is a variable in Y for some x. A bounding of a free variable is a defining the function and that is computable if this operation carrying out under the β-reduction.  That reduction of λ-calculus is recursive if the bounding and substituting are consistent. The rule of λ-calculus as follows:
If z is not a free variable of Y, then: (λx.Y)z ⊢ Y[x:=z];
If x is bounded by Y and Y don't contains any free variable in z, then (λx.Y)z = Y[x:=z], that yielded function Y = f(z).

That means the λ–calculus is a kind of computability by human's abstract operations
The λ-calculus can be bounding in itself (recursive):

λ f
λ f.x
λ f.f(x)
λ f.f(f(x))
λ f.f(f(x))
…

That is corresponding nature numbers: 0, 1, 2, 3, 4….

Any application of λ-calculus acts on a free variable of the application itself that will result unpredictable consequence, namely if it contains more one sub- λ form of itself, the application is nondeterministic, like this definition:

λ ((λx. f(y)). (λx. f(y)).

This definition is simultaneous (synchro) form of two abstracted operations, but which occur within the coinstantaneous time, and then two abstractions can't separate in actual time.

### 4. Nondeterministic: DTM,NTM and TM

The prevalent definition about *nondeterministic* originated from the *Nondeterministic TM* (NTM), the prevalent concept of NTM is not original from TM rather automaton theory —the nondeterministic automaton, the *nondeterministic* and *deterministic* are connatural in the light of technostructure:

" when the machine is in a given state and reads the next input symbol, we know what the next state will be —it is determined. We call this deterministic computation. In a nondeterministic machine, several choices may exist for the next state at any point. Nondeterminism is a generalization of determinism, so every deterministic finite automaton is automatically a nondeterministic finite automaton." (M. Sipser[9] section1.2, p.47)

"The formal definition of a nondeterministic finite automaton is similar to that of a deterministic finite automaton. Both have states, and input alphabet, a transition function, a start static, and a collection of accept states. However, they differ in one essential way: in the type of transition function. In a DFA the transition function takes a state and an input symbol and produces the next state. In an NFA the transition functions takes a state and an input symbol *or the empty string* and produce *the set of possible next state*." (M. Sipser[9] section1.2, p.53)

These definitions based on the internal state, the difference between Deterministic TM (DTM) and Nondeterministic TM (NTM) are the formation of internal state, which is distinguished from the notion of TM. In fact people have unwittingly changed the notion of TM, and the prevalent definition of NTM is defined similar with the DTM rather original TM:

"We now define a Turing machine to be a finite set of quadruples, no two of which begin with the same pair qisj. Actually, any finite set of quadruples is called a nondeterministic Turing machine. But for the present we will deal only with deterministic Turing machines which satisfy the additional 'con-sistency' condition forbidding two quadruples of a given machine to begin with the same pair qisj, thereby guaranteeing that at any stage a Turing machine is capable of only one action…. We now define a Turing machine to be a finite set of quadruples, no two of which begin with the same pair qisj. Actually, any finite set of quadruples is called a nondeterministic Turing machine. But for the present we will deal only with deterministic Turing machines which satisfy the additional 'con-sistency' condition forbidding two quadruples of a given machine to begin with the same pair qisj, thereby guaranteeing that at any stage a Turing machine is capable of only one action." (M. D. Davis and E. J. Weyuke [8])

 Apparently the NTM and DTM are computable in homogeneous (*con-sistency* or single) internal state, the difference between them is only the formation of configuration. It should be noted that this definition of DTM distinguishes from "TM is Decidable" (above section 1. Two key points), DTM and NTM are about computable problem but the TM for decision is judgment problem. In another word the TM is just DTM when they are about computable problem; but this meaning differs from the TM for decidable, the latter is relative with the *Undecidable*.
 The common definition NP problem is based on this definition of NTM, the prevalence two definitions of NP problem, one of them that is solvable by NTM, just derived from this notion.
 But the concept of *Nondeterministic* is notated else the *Nondeterministic Polynomial Time*, that is to say the configuration of internal state has replaced with the formation of polynomial time, which implies the actual time of TM has been replaced with the form of polynomial time, this arrogate became the misdirection of the DTM became the decision TM:

"**DEFINITION 7.16** NP is the class of languages that have polynomial time verifiers. …The term NP comes from nondeterministic polynomial time and is derived from an alternative characterization by using nondeterministic polynomial time Turing machines.
**THEOREM 7.17** A language is in NP iff it is decided by some nondeterministic polynomial time Turing machine.
**PROOF IDEA** We show how to convert a polynomial time verifier to an equivalent polynomial time NTM and vice versa. The NTM simulates the verifier by guessing the certificate. The verifier simulates the NTM by using the accepting branch as the certificate.
**DEFINITION 7.9** Let *N* be a nondeterministic Turing machine that is a decider. The *running time* of *N* is the function f: N→N, where f(n) is the maximum number of steps that *N* uses on any branch of its computation on any input of length n, … The definition of the running time of a nondeterministic Turing machine is not intended to correspond to any real-world computing device. Rather, it is a useful mathematical definition that assists in characterizing the complexity of an important class of computation problems, as we demodtrate shortly.
**THEOREM 7.10** let t(n) be a function, where t(n) ≥n. The every t(n) time nondeterministic single-tape Turing machine has an equivalent $2^{O(t(n))}$ time deterministic single tape Turing machine.
… we defined the time of a nondeterministic machine to be the time used by longest computation branch." (M. Sipser [9])

The *longest computation branch* here is still a polynomial time, so called the *nondeterministic polynomial time* is but a pending state upon some polynomial-times branches. This stochastic choice in internal state is no more than replacing the Turing's *oracle* with a kind of *Guess*, therefore the DTM became a special case of NTM, namely P ∈ NP, in this circumstance, one may constrainedly adduce the Church-Turing Thesis: P=NP, namely they are equivalent in computability. Thereby this definition of NTM is pointless, this way leads an inkling of the NP may be something else.

  In fact the properties of transition functions are disparate essentially among the DTM, NTM and TM, for the former of DTM and NTM the both are similar, but the latter of TM is different essentially from former: either the recursive internal state (computable) or non-recursive (Undecidable).

  The non-recursive internal state is the truer definition of the *Nondeterministic* namely the *Nondeterministic* is just the Undecidable Decision of computation. This definition of *Nondeterministic* is relative to the Computability, contrarily, the prevalent definition of NTM with the alternative *n-tuples* is "incomputable" (the oracle), because the stochastic choice in internal state is not satisfiable in actual time coinstantaneously.

  Although the property of internal state is the essential connotation of the computer, but the prevalent sense about TM and automaton are based on the space capability (the long of tape or amount of storage). People often think the Turing Machines is equivalent with automaton but has an unlimited amount of storage space, in fact the difference between TM and automaton based on with or without the internal state; The difference between TM and NTM is based on the property of internal state.

**5. The P-reduction and Isomorphism**

  Although the NP (Nondeterministic Polynomial Time) was defined equivalence with NTM, the notation NP in the concepts *NP problem* and *NP-completeness Problem* has become another great trouble that has been arising from the TUM as the Turing Reducible with its oracle.

  Stephen A. Cook, in his paper in 1971 "*The Complexity of Theorem-Proving Procedures*" [4] introduced the Polynomial-reduction in his compound machine imitated the Turing's UM:

"Here 'reduced' means, roughly speaking, that the first problem can be solved deterministically in polynomial time provided an oracle is available for solving the second."

"A query machine is a multitape Turing machine with a distinguished tape called the query tape, and three distinguished states called the query state, yes state, and no state, respectively. If M is a query machine and T is a set of strings, then a T –computation of M is a computation of M in which initially M is in the initial state and has an input string w on its input tape, and each time M assumes the query state there is a string u on the query tape, and the next state M assumes is the yes state if u ∈ T and the no state if u ∉ T. We think of an 'oracle', which knows T, placing M in the yes state or no state.
Definition. A set S of strings is P-reducible (P for polynomial) to a set T of strings iff there is some query machine M and a polynomial Q (n) such that for each input string w, the T-computation of M with input w halts within Q(|w|) steps (|w| is the length of w) and ends in an accepting state iff w ∈ S."

   Apparently there is a decidable judgment by a compound mechanism that consists of the T-computation and the query means, the T-computation of M accepts the input string, in another hand M act as judgeship to check the state of T-computation. In fact there are two entangling hierarchy of computations (the T-computation (of M) and the M is the relation —Halting or accepted) and logics (checking and deciding —ended or stop). The former is the P-reduction as computation, the latter is a logic judgment about set.
   In Cook's M, The T –computation is whole configuration of computer and the M is also decidable in checking, Cook specially defined the T-computation in the appendix of his paper "The P versus NP Problem" (2000):

"The computation of M on input w ∈ Σ* is the unique sequence $C_0, C_1,...$ of configurations such that $C_0 = q_0w$ (or $C_0 = q_0b$ if w is empty) and $C_i \to C_{i+1}$ for each i with $C_{i+1}$ in the computation, and either the sequence is infinite or it ends in a halting configuration. If the computation is finite, then the number of steps is one less than the number of configurations; otherwise the number of steps is infinite. We say that M accepts w iff the computation is finite and the final configuration contains the state q accept. "

   The M chicks the T-computation on the taper instead of the state, this definition is very different from TUM that is reconstructed in intention state in same machine, but Cook's M and its T-computation are relation between two computation with the *query tape* that is the both as the otput of T-computation and input of the query machine M, namely, the T-computation accepts the strings and M decides the result of T-computations.
Directly say, the TUM is untenable for decidable judgment (*Undecidable Decision*) but the Cook's is the compound machine and unfathomable P-reducible. The Turing's corollary is that the Hating problem is undecidable, the Cook's is the NP-completeness. If accepting this definition that means we have admitted this replacing of oracle with the P-reducible. In fact whether the Cook's P-reducible is consistent with Turing Reducible that is still not to be debated adequately, the definition of P-reducible has used in consequent proving.
   Cook brought forward two new notions: the *subgraph problem* and the *graph isomorphism* problem attempted to redefine the P-reducible in logic tautologies:

"1. The subgraph problem is the problem given two finite undirected graphs, determine whether the first is isomorphic to a subgraph of the second. A graph G can be represented by a string ⁻G on the alphabet {0,1,*} by listing the successive rows of its adjacency matrix, separated by *s. We let {subgraph pairs} denote the set of strings ⁻G1**⁻G2 such that G1 is isomorphic to a subgraph of G2.
2. The graph isomorphism problem will be represented by the set, denoted by {isomorphic graphpairs}, of all strings ⁻G1**⁻G2 such that G1 is isomorphic to G2.
3. The set {Primes} is the set of all binary notations for prime numbers.
4. The set {DNF tautologies} is the set of strings representing tautologies in disjunctive normal form.
5. The set D3 consists of those tautologies in disjunctive normal form in which each disjunct has at most three conjuncts (each of which is an atom or negation of an atom).

**Theorem 1.** If a set S of strings is accepted by some nondeterministic Turing machine within polynomial time, then S is P-reducible to {DNF tautologies}.

**Corollary.** Each of the sets in definitions 1)–5) is P-reducible to {DNF tautologies}.

This is because each set, or its complement, is accepted in polynomial time by some nondeterministic Turing machine.

Proof of the theorem. Suppose a nondeterministic Turing machine M accepts a set S of strings within time Q(n), where Q(n) is a polynomial. Given an input w for M, we will construct a proposition formula A(w) in conjunctive normal form such that A(w) is satisfiable iff M accepts w. Thus ¬A(w) is easily put in disjunctive normal form (using De Morgan's laws), and ¬A(w) is a tautology if and only if w ∉ S. Since the whole construction can be carried out in time bounded by a polynomial in |w| (the length of w), the theorem will be proved. "

It should be noted that there are two kinds of sets: the graph ⁻G ({0,1,*}), *the subgraph pairs* (⁻G1**⁻G2) and the {subgraph pairs} (the set of strings ⁻G1**⁻G2) that as the subgraph proglems are enumerable reclusive because they are sequences and there always the alphabetical order; but *the* {isomorphic graphpairs} (all strings ⁻G1**⁻G2) as the isomorphic problem is innumerable. Cook notarized this point:

" **Theorem 2.** The following sets are P-reducible to each other in pairs (and hence each has the same polynomial degree of difficulty): {tautologies}, {DNF tautologies}, D3, {subgraph pairs}.

Remark. We have not been able to add either {primes} or {isomorphic graphpairs} to the above list. To show {tautologies} is P-reducible to {primes} would seem to require some deep results in number theory, while showing {tautologies} is P-reducible to {isomorphic graphpairs} would probably upset a conjecture of Corneil's from which he deduces that the graph isomorphism problem can be solved in polynomial time."

We may make out two different isomorph problems:
1. The subgraph problem: {subgraph pairs} (⁻G1**⁻G2), set {DNF tautologies}, —decidable;
2. The graph isomorphism problem: {isomorphic graphpairs}, —undecidable.

The isomorph problems is virtually constructing new relative strings in a set (for instance the graph ⁻G by *s) or between sets (the subgraph pairs), the {subgraph pairs} by**), the P-reducible is defined only the latter, simply denotes ** ≤ p  **; but practically between the former and the latter namely between two isomorph, simply denotes * ≤ p ** ≤ p  {**}.

Apparently the definitions 1)–5) are not equipollent each other, but the Corollary of Theorem 1 clamed: Each of the sets in definitions 1)–5) is P-reducible to {DNF tautologies}). This proof only resorted to the prevalent questionable definition of NTM and the {DNF tautologies} acted as dual role.

   The prevalent concept NTM has been defined a TM in the state upon some computations deterministic branches (as above mentioed), thereby the isomorph in one set or between sets has the same polynomial degree of facility (that should allow invoking the Church-Turing Thesis), the P-reducible here is but a transformer (reconstruction) between strings or between sequences, thereby the Theorem 1 is evident, the set {DNF tautologies} is paratactic disjunctions that equivalence with NTM, we denotes "**", correspondingly the tautologyhood is completive form namely the CNF, denotes "{**}". That is to say the {DNF tautologies} is virtually  Boolean form (implication), but the {DNF tautologies} as logic tautologie is acquiescent reducible to the tautologyhood (CNF, virtually logic tatology), thereby the P-reducible is practically the relations ** ≤ p  {**} (the set {DNF tautologies} is reduced to the tautologyhood CNF). The tautology as logic form is always true in every possible, the completeness of logic form is the strict implication (that could be decided instantly by oracle); by the  p  ** ≤ p  {**}, the {DNF tautologies} as Boolean form and possessed the Satisfiability become  logic tautology (strict implication), and the P-reducible became a material implication (paradox).

   The {DNF tautologies} acted as dual role of tautology (DNF that *can be reduced to determining tautologyhood*) and tautologyhood (*could be decided instantly (by an oracle)*):

"The set of tautologies (denoted by {tautologies}) is a certain recursive set of strings on this alphabet, and we are interested in the problem of finding a good lower bound on its possible recognition times. We provide no such lower bound here, but theorem 1 will give evidence that {tautologies} is a difficult set to recognize, since many apparently difficult problems can be reduced to determining tautologyhood. By reduced we mean, roughly speaking, that if tautologyhood could be decided instantly (by an oracle) then these problems could be decided in polynomial time." (S. Cook 1971)

   That is to say, the DNF is both the recursive set (determining tautologyhood) and undecidable set (have not lower bound recognition times namely the *tautologyhood could be decided instantly (by an oracle)*). Duo to the questionable P-reducible the Satisfiability of Boolean form became the completeness of tautology (oracle). Therefore the subgraph problem became the graph isomorphism problem; the P-reducible interweaved in isomorphic, which originates all trouble problems in NP theory.

## 6. The P-reduction and Semi-decidable

The P-reducible has re-explained as a binary relation on set — the semi-decidable, which is the origin of another prevalence definition of NP problem:

"The notion NP stands for 'nondeterministic polynomial time', since originally NP was defined in terms of nondeterministic machines (that is, machines that have more than one possible move from a given configuration). However, now it is customary to give an equivalent definition using the notion of a checking relation, which is simply a binary relation R ⊆ Σ* × Σ * 1 for some finite alphabets Σ and Σ1. We associate with each such relation R a language LR overΣ1 ∪ Σ1 ∪ {#} defined by LR = {w#y| R(w, y)} where the symbol # is not in Σ. We say that R is polynomial-time iff LR ∈ P.
  Now we define the class NP of languages by the condition that a language L over Σ is in NP iff there is k ∈ N and a polynomial-time checking relation R such that for all w ∈ Σ, w ∈⇔ ∃y(|y| ≤ |w R(w, |k and y)), where |w| and |y| denote the lengths of w and y, respectively. "(S. Cook, 2000 [6])

The LR = {w#y| R(w, y)} ∈ P, then the R ⊆ Σ* × Σ * is polynomial-time and is equivalent with the P-reducible, which is equipollent with the {subgraph pairs} (¯G1**¯G2). Cook thinks the semi-decidable, P-reducible, even HP is computably enumerable:

"The computability precursors of the classes P and NP are the classes of decidable and c.e. (computably enumerable) languages, respectively. We say that a language L is c.e. (or semi-decidable) iff L = L(M) for some Turing machine M. We say that L is decidable iff L = L(M) for some Turing machine M that satisfies the condition
That M halts on all input strings w. There is an equivalent definition of c.e. that brings out its analogy with NP, namely L is c.e. iff there is a computable 'checking relation ' R(x, y) such that L = {x |∃y R(x, y)}. Using the notation <M> to denote a string describing a Turing machine M, we define the Halting Problem HP as follows:
HP={ <M> |M is a Turing machine that halts on input <M> }.
Turing used a simple diagonal argument to show that HP is not decidable. On the other hand, it is not hard to show that HP is c.e." (S. Cook 2000: 2.History and Importance)

That is to say the HP here is Turing' J rather than J'[2] (§8), which contradicted to Turing's proving. This binary relation R ⊆ Σ* × Σ * 1 may be computably enumerable, but the LR = {w#y| R(w, y)} as a set is an Undecidable Decision, because of the set of language {w|over Σ } = Σ* is undecidable, that is the same with {isomorphic graphpairs}.
  Using the notation <R > to denote the string describing (w#y)##( w#y), the R for the <R> is undecidable.

**7. The Oracle and NP-completeness**

The Turing Reducible with the oracle now has defined for language formally: L 1 is Turing Reducible to L 2, if L 1 can be decided by a UM using L 2 as its oracle. The oracle here should been apprehended omniscient logic or omnipotent algorithm, the oracle for Turing was perhaps an opposite of the computability to avoid using like such word of incomputability that is a semantics paradox (but this usage distinguishes from the notion of incomputable), in fact the oracle is essentially is the meaning of omnipresent and should been comprehended the actual time.

According this comprehension, the Turing Reducible with its oracle is just the *Undecidable Decision* in actual time rather than the decidable in actual time. Martin Davis in his paper "what is Turing Reducibility?" thought:

"The concept of Turing reducibility has to do with the question: can one non-computable set be more non-computable than another? In a rather incidental aside to the main topic of Alan Turing's doctoral dissertation (the subject of Soloman Feferman's article in this issue of the Notices), he introduced the idea of a computation with respect to an oracle. An oracle for a particular set of natural numbers may be visualized as a 'black box' that will correctly answer questions about whether specific numbers belong to that set. We can then imagine an oracle algorithm whose operations can be interrupted to query such an oracle with its further progress dependent on the reply obtained. Then for sets A,B of natural numbers, A is said to be Turing Reducible to B if there is an oracle algorithm for testing membership in A having full recourse to an oracle for B. The notation used is: A ≤t B. Of course, if B is itself a computable set, then nothing new happens; in such a case A ≤t B just means that A is computable. But if B is non-computable, then interesting things happen." [7]

That is to say the Turing Reducible is either repeating The Church-Turing Thesis (that has the same polynomial degree of facility rather than difficulty) otherwise undecidable, in fact the degree of difficulty as computation comparability is untenable. As the former the Turing Reducible is but a computable transition function —a transformer of same degree of facility, but for second circumstances just the Turing's undecidable. Turing's oracle misdirects later the definition of reducible into great troubles.

In NP Theory, people often use the method of re-redefining a definition to substitute the inference process, that is a kind of *Begging the Question* or *Material Fallacies* so called, As Martin Davis said: "I am thinking of theoretical computer science particularly, where the field moves very fast, leaving unsolved problems behind, under the assumption that, because they haven't been able to deal with the problems in a year or two, the problems are intractable. And they go on to another subject. Following the habits in the more applied parts of computer science, the intellectual center is not finished public articles, but rather conferences. A conference program committee—I have been on one of them —get extended abstracts in which theorems are stated but rarely proved, and somehow judgments are made. But the things that are claimed to be proved are not always in fact proved. And the stakes are high." [11])

The re-defining is a pure research method in theory differs from induction and analysis, with which the reduction (may be called the deductive-reduction) is similar, but lakes strictness unless that will be guaranteed by proving, for example the λ-calculus could be regarded as deductive-reduction guaranteed by the β-reduction strictly.
   In the NP theory, the notation NP is virtually the opposability of computability, that should be abbreviation of *Nondeterministic Problem*, because for that there is no a deterministic algorithm, the *oracle* (omniscience) of Turing Reducible is instantaneous that is transcendental to the *actual time* (as the essence of algorithm). But the P-*reducible* alters unwittingly from computable checking to set comparison of a kind, via which the Turing Reducible is re-redefined the *Polynomial Re-Reducibility* that is equipollent with the binary of computably enumerable or semi-decidable:

"We say that a language L is c.e. (or semi-decidable) iff L = L(M) for some Turing machine M. We say that L is decidable iff L = L(M) for some Turing machine M that satisfies the condition that M halts on all input strings w. There is an equivalent definition of c.e. that brings out its analogy with NP, namely L is c.e. iff there is a computable 'checking relation' R(x,y) such that L = {x|∃yR(x,y)}." (S. Cook 2000: 2.History and Importance)

   That is to say, there is dual relation of P-reduction form, which is both the checking and synchronously deciding, that corresponding isomorph is to construct transcendentally a P-reduction between subgraph problem and graph isomorphism problem. Namely, the oracle (omniscience) in Turing's Reducible is instantaneous (omnipotent) relation in P-reducible, the computable checking in polynomial time becomes the decidable instantaneously. Furthermore, the binary of computably enumerable or semi-decidable became the property of completeness namely *c.e.-completeness* and the *NP-completeness*:

 "The notion of NP-completeis based on the following notion from computability theory:
**Definition2.** A language L is  c.e.-complete iff L is c.e., and L ≤m L for every  c.e. languageL."
"NP-completeness as a polynomial-time analog of c.e.-completeness, except that the reduction used was a polynomial-time analog of Turing reducibility rather than of many-one reducibility. The main results are that several natural problems, including Satisfiability and 3-SAT  (defined below) and subgraph isomorphism are NP-complete. A year later Karp used these completeness results to show that 20 other natural problems are NP-complete, thus forcefully demonstrating the importance of the subject. Karp also introduced the now standard notation P and NP and redefined NP-completeness using the polynomial-time analog of many-one reducibility, a definition that has become standard. Meanwhile Levin, independently of Cook and Karp, defined the notion of 'universal search problem', similar to the NP-complete problem, and gave six examples, including Satisfiability.
The standard definitions concerning NP-completeness are close analogs of Definitions 1 and 2 above.

**Definition3.** Suppose that Li is a language over Σ, i=1,2. Then L 1≤p L2 (L1 is p-reducible to L2) iff there is a polynomial-time computable function f: Σ →Σ such that x ∈ L1 ⟺ f(x) ∈ L2, for all x ∈ Σ*1.

**Definition4.** A language L is NP-complete iff L is in NP, and L'≤p L for everyLanguage L' in NP."

"the more restricted notion of many-one reducibility (≤m) was introduced and defined as follows.

**Definition1.** Suppose that Li is a language over Σ, i=1,2. Then L1≤m L2 iff there is a (total) computable function f: Σ*1→Σ*2 such that x ∈ L1 ⟺ f(x) ∈ L2, For all x ∈ Σ*1.

It is easy to see that if L1 ≤ m L2 and L2 is decidable, then L 1 is decidable.

This fact provides an important tool for showing undecidability; for example, if HP ≤m L, then L is undecidable.

The notion of NP-complete is based on the following notion from computability theory:

**Definition2.** A language L is c.e.-complete iff L is c.e., and L ≤m L for every c.e. language L'.

It is easy to show tha HP is c.e.-complete." (S. Cook 2000: 2.History and Importance)

The reduction via isomorph, checking relation, semi-decidable, made the NP problem to be P-reduced to the NP-completeness, even the famous HP. the P-reducible as a transformer of same degree of facility has been reduced a judgment in the form of logic formal implication and gains the formal completeness.

This notion of NP-completeness is intensified by polynomial-bounded algorithms by R. M .Karp:

"It is reasonable to consider such a problem satisfactory solved when an algorithm for its solution is found which terminates within a number of steps bound by a polynomial in the length of the input. We show that a larger number of classic unsolved problem of covering, matching, packing, routing, assignment and sequencing are equivalent, in the sense that either each of them possesses a polynomial-bounded algorithm or none of them does." "We are specifically interested in the existence of algorithms that are guaranteed to terminate in a number of steps bounded by a plynomia in the length of the input. We exhibit a class of well-known combinatorial problem, including those mentioned above, which are equivalent, in the sense that a polynomial-bounded algorithm for any one of them would effectively yield a polynomial-bounded algorithm for all. We also show that, if these problem do possessed polynomial-bounded algorithm then all the problems in an unexpectedly wide class (roughly speaking, the class of problems solvable by polynomial-depth backtrack search) possess polynomial-bounded algorithms." (R. M .Karp 1972) [5]

So called *polynomial-bounded algorithm* is but another name of computable checking relation, that is only the polynomial checking relation to be used in the searching algorithm, for example the *polynomial-depth backtrack search* is a P-reducible to approximation value, namely that is a concrete P-reducible method in optimal algorithm, roughly the checking relation served as algorithm, in this meaning the completeness here is the an algorithm-completeness rather than the so called NP-completeness.

After simplified the prevalently proof of Cook-Levin Theorem (Boolean satisfiability problem is the NP-completeness) as follows:
 "There are two parts to proving that the Boolean satisfiability problem (SAT) is NP-complete. One is to show that SAT is an NP problem. The other is to show that every NP problem can be reduced to an instance of a SAT problem by a polynomial-time many-one reduction."(Quoted form Wikipedia)

   In fact, these proving draw support from the questionable prevalent definition of NTM and the debatable P-reducible, which has mentioned above. Perhaps the Cook–Levin theorem had proved an equivalence of two uncertain concepts, in fact this theorem hinds two preconditions, "if the definition of P-reducible is tenable, …" and "the Boolean form is satisfiability", and the two preconditions themselves are unconfirmed concerts thereby this theorem is double nondeterministic; The Cook-Levin theorem is an equipollent of two nondeterministic.
The first part is apparent according to the prevalent definition of NTM, the other according to the P-reducible is easy, this proving seemed over easy, the reducible acted as all-rounder transformer.
   The Boolean form as the implication is equivalent with the prevalent definition of NTM that is also equivalent with the NP-problem, whether they can be reduce to the SAT as the logic tautology that is unfathomed, simply, the satisfiability of Boolean form is conditionally a logic implication, and the completeness is simply the logic tautology, the problem "Boolean formula is satisfiable" distinguishes essentially form the completeness of logical form (oracle), this proof is virtually material implication paradox.
 The title "The Complexity of Theorem-Proving Procedures" and the main section "Tautologies and Polynomial Re-Reducibility" in Cook's paper seemed to show these difficulty.
   Of course all these argument are still significance for in-depth research in the NP theory, Hao Wang (1921-1995) thought the CNF is the most important problem in mathematical logic, perhaps that "To be is to be a value of a variable" (W.wan O. Quine [10],) is the best portrayal for CNF,  (that means we only commit ourselves to an ontology by claims that say things like 'There is something (bound variable) that is …'
For example, 'There is something that is a prime number greater than a million' commits us to believing that such numbers are entities. And 'There is some property (or characteristic) that red houses and red cars have in common' commits us to believing that properties (or characteristics) are entities. Quine's slogan [14] doesn't tell us which ontology is true or which we should accept. It only tells us how we commit ourselves to a given ontology.)

**8.  The NP Theory: NP ≠ P**

There should be no problem if the definitions of P and NP are respectively isolative or identical, for example P=NP according to the prevalent definition of NTM (section 4. above), the trouble comes from the relation of "P verses NP problem", which is root in the relation of TM with the questionable prevalent concept NTM:

"The P versus NP problem is to determine whether every language accepted by some nondeterministic algorithm in polynomial time is also accepted by some (deterministic) algorithm in polynomial time." (S. Cook 2000)

This proposition roughly is like this: *whether NTM = TM for language*.
The remarkable notion *nondeterministic algorithm* and its relative *deterministic algorithm* are just the *nondeterministic polynomial time* and (*deterministic*) polynomial time, which are similar to Karp's *polynomial-bounded algorithms*, they all are homogenous oriented from the misdirect dimorphism of TM with NTM.
This problem have analyzed as above. the prevalent definition of NP problem (or prevalent NTM) is ultimately the *deterministic algorithm* or polynomial time algorithm, the difference is the *con-sistency* from single in internal state, then simply NTM=TM in essence. But "the P versus NP problem" descripted *whether NTM = TM (for language)*, simply, NTM ≠TM or strictly NTM ≠ DTM (Deterministic TM).

(the *nondeterministic algorithm* in polynomial time is not equivalent with the *deterministic algorithm* in polynomial time).

The real question concerns the relation of computable number (nature number) with language.

Based on the investigation above, there are arguments for the P versus NP problem:
Every language connotes all languages that distinguish from the some languages, simply, the former is the Undecidable Decision. Some language accepted by some NTM in polynomial time is P problem, or speaking they are all equivalent in degree of difficult; according to the definition of NP (Nondeterministic Problem) the nondeterministic problems all are NP-completeness, the P versus NP problem is virtually the uncertain NP-completeness. This situation redefines the NP problem and P-reducible together in interweaving hiberarchy again, in which the NTM with the polynomial time acted as the P-reduction in DTM, this intractability is just the unreadable NP problem.

For more strictly discussing these problems, there are two pairs of concepts to be introduced in NP theory:
1. P-problem (the definition of decidable problem)
2. P-algorithm (effective computable algorithm)[12], the both are identical (may be called P-completeness that means algorithm is completeness for problems).
3. There is nondeterministic relations between NP (Nondeterministic Problem rather than Nondeterministic Polynomial Time)

4. NP algorithm (defines on the optimal algorithm to get the best fit approximation value for NP). There is the true problem how the judgment to NP forms a decision by a NP-algorithm?

According to previous definition *nondeterministic polynomial time*, there is no necessary of definition of NP-algorithm, also no kind of NTM-algorithm; in fact the true definition of *nondeterministic problem* (NP) necessitates, which is just the cause for arising of many methods of heuristic algorithms. For instance, the *polynomial-depth backtrack search* method by R. M .Karp used is just the most basic NP-algorithm. In fact as NP-algorithm there are larger numbers of classic unsolved problems in real life, which are corresponding the true definition of NP and NP-algorithm.

Altogether, in the viewpoints above all, the conclusions are as follows:
1. The P-reducible is misdirected from the Turing Reducible with its oracle.
   The NP-completeness is a reversal to The Church-Turing Thesis;
   The Cook-Levin theorem is an equipollent of two uncertain.
2. According to the previous definition of NTM and NP, P ∈ NP, and adduce constrainedly the Church-Turing Thesis, P=NP (under this definitions). Therefore the P versus NP problem is meaningless.
3. According the notion of the *Nondeterministic Problem* and NP-algorithm, NP is the *Undecidable Decision*, P≠NP, because the two are disparate.
4. Correctly, P versus NP is computability versus Nondeterministic.
5. The NP-algorithm is effective approximate way by TM.
6. The deterministic algorithm or computability between P and NP is NP-hard.

**Discussion**:

1. The development of NP theory has had most uneven history, which in more than half a century, established a new landmark in academic and IT industry and attracted thousands on thousands of people join in it, although there are serious difficulties, the NP theory profoundly enlightens the relations computability, algorithm and computer theory. By the contributions of many scholars has made and will make the NP theory should be a greater discipline among the Computer science, Math-Logic, Information technology, Artificial Intelligence, Engineering science, Mathematics even Culture and Philosophy.
2. The NP theory with the actual time is enlighten between theory and reality, the NP is an emersion of classical contradiction of the potential and real infinite in philosophy, for example the linear actual time as the polynomial time differs essentially from the unfathomable "exponential time", that even involves the Continuum Hypothesis (the continuum has the same cardinality as the power set of the integers), which may regard a reappearance of Godel's incompleteness theorem.
3. The mathematics is necessarily Hilbert's formalism but a mathematician is better Brower's intuitionist.

**Acknowledgement**: Thanks to Professor Yu LI (Université de Picardie Jules Verne) for her support, communication and interviews to me.

itself.  All about connotations of these concepts that involves about relations of mathematics and logic will not be furthermore investigated here.